# The influence of size effect on the electronic and elastic properties of diamond films with nanometer thickness

*Leonid A. Chernozatonskii[†], Pavel B. Sorokin[†,‡,§], Alexander A. Kuzubov[‡], Boris P. Sorokin[#], Alexander G. Kvashnin[‡], Dmitry G. Kvashnin[‡], Pavel V. Avramov[‡,††,⊥] and Boris I. Yakobson[§]*

Emanuel Institute of Biochemical Physics, Russian Academy of Sciences, 4 Kosigina st., Moscow, 119334, Russian Federation

Siberian Federal University, 79 Svobodny av., Krasnoyarsk, 660041 Russian Federation

Department of Mechanical Engineering & Material Science and Department of Chemistry, Rice University, Houston, Texas 77251, USA

Technological Institute of Superhard and Novel Carbon Materials, 7a Centralnaya Street, Troitsk, Moscow region, 142190, Russia

Kirensky Institute of Physics, Russian Academy of Sciences, Akademgorodok, Krasnoyarsk, 660036 Russian Federation

Advanced Science Research Center, Japan Atomic Energy Agency, Tokai, Ibaraki319-1195, Japan

[†] Emanuel Institute of Biochemical Physics
[‡] Siberian Federal University
[§] Rice University
[#] Technological Institute of Superhard and Novel Carbon Materials
[††] Kirensky Institute of Physics
[⊥] Advanced Science Research Center

cherno@sky.chph.ras.ru

Corresponding author. E.mail: cherno@sky.chph.ras.ru

The atomic structure and physical properties of few-layered $\langle 111 \rangle$ oriented diamond nanocrystals (diamanes), covered by hydrogen atoms from both sides are studied using electronic band structure calculations. It was shown that energy stability linear increases upon increasing of the thickness of proposed structures. All 2D carbon films display direct dielectric band gaps with nonlinear quantum confinement response upon the thickness. Elastic properties of diamanes reveal complex dependence upon increasing of the number of $\langle 111 \rangle$ layers. All theoretical results were compared with available experimental data.

*Introduction*

The graphene as two-dimensional material has attracted attention from the scientific community long before experimental fabrication. The first theoretical study of graphene is dated 1946, when the remarkable band structure of "Dirac cones" has been studied [1]. First experimental observation of free standing graphene [2] in 2004 initiated the comprehensive study of this material. Ballistic conductivity, pseudo-chiral Dirac's nature of carriers, anomalous Hall effect [3] makes graphene the most promising material for science and future technology.

Hydrogenation of graphene enlarges its potential application in nanoelectronics. Regular adsorption of hydrogen atoms changes graphene electronic structure and opens the band gap depending upon the distance between hydrogen regions [4-8]. Total hydrogenation of graphene changes the nature of electronic states due to changing of $sp^2$ hybridization of C-C bonds to $sp^3$ one and opens the dielectric band gap. [9,10] Such two-dimensional insulator was called as graphane. [10] The theoretical prediction was generally confirmed experimentally by Elias *et. al* [11].

Graphane is an offspring of graphene along with graphene nanoribbons and carbon nanotubes. The other type of carbon bonding opens a new way for developing of two-dimensional carbon based materials.

Graphane is the first member in a series of $sp^3$ bonded diamond films consist of a number of adjusted $\langle 111 \rangle$ oriented layers which display unique physical properties. For the first time diamane structures were proposed by Chernozatonskii et. al. [12] in 2009. Recently [13] similar $C_2H$ structure was also considered.

Usually diamond films prepared by CVD method with micrometer thickness. [14] Also the diamond quantum wells obtained in the bulk structures of superlattices. [15] The consequent study of graphene, graphane and proposed diamanes can be considered as bottom-up nanotechnological approach opposite to ordinary top-down paradigm. The main goal of this work is to study diamane physical properties. As against to Ref. 12 we consider diamanes with different thickness, we investigate their stability and compare them with known data for sp$^3$-hybridized hydrocarbon clusters. We study the elastic properties of the structures and obtain phonon dispersion, wave velocities and elastic constants. Finally we discuss possible ways to synthesize the structures.

*Method and model*

The plane wave DFT PBE [16] electronic structure calculations of 2D carbon nanostructures were performed using Ultrasoft Vanderbilt pseudopotentials [17] and a plane-wave energy cutoff equal to 30 Ry by PWSCF code [18]. To calculate equilibrium atomic structures, the Brillouin zone was sampled according to the Monkhorst–Pack [19] scheme with a 16×16×1 k-point convergence grid. To avoid interactions between the species, neighboring planes were separated at least by 10 Å in the hexagonal supercells. During the atomic structure minimization, structural relaxation was performed until the change in total energy was less than $3\times10^{-7}$ eV. Phonons calculations were performed within density functional perturbation theory. [20] All the values given above were carefully tested and found optimal.

*Results and discussion*

Atomic structures of graphene and graphane are presented in Fig. 1a and 1b, respectively. Both structures display hexagonal symmetry with essentially different lattice parameters (2.468 and 2.540 Å, respectively) because of different nature of chemical bonding. In contrast to flat $sp^2$-hybridized graphene, crimped $sp^3$ graphane is characterized by two terminal layers of hydrogen atoms from both sides of the sheet. Surface hydrogen atoms of graphane at least from one side can be replaced by one more layer of $sp^3$ carbon atoms forming a diamane structure. Since chemical bonding between carbon layers in diamanes is also realized by $sp^3$ hybridization, the lattice parameters of studied diamanes are close to each other ($a = 2.53$ Å).

Hydration of graphene completely changes the nature of the material which can be considered as the thinnest possible diamond. The diamond films with smallest thickness obviously succeed the properties of the graphane with consistent approaching to the bulk diamond limit. Diamond films or diamanes under study fill the gap between two-dimensional graphane and crystalline diamond.

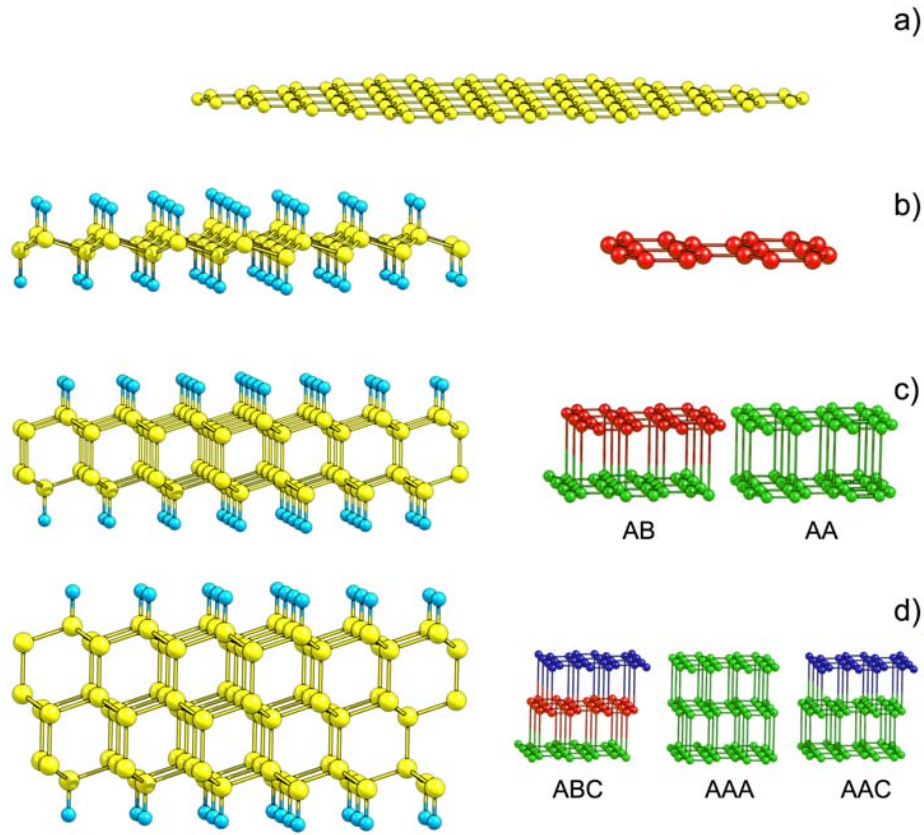

**Fig. 1. The atomic geometry of studied quasi two-dimensional carbon nanofilms a) graphene; b) graphane; c) diamane D(AB) with AB stacking of two layers; d) diamane D(ABC) with ABC stacking of three layers. Carbon atoms are marked by yellow (light gray), hydrogen are marked by blue (gray). For each diamane possible stacking sequences of carbon layers are presented.**

The atomic geometry of the diamane consists of stacking covalently bounded monoatomic carbon layers (Fig. 1b-d). The changing of the stacking sequence allows constructing different diamane polytypes. According to this fact diamanes can be compactly classified as D(ijk…l) where i, j, k, l are positions layer types which can be equal to A, B or C. For example, two and three layered diamanes with natural diamond stacking sequence (ABCABC…) are denoted as D(AB) and D(ABC); two and three layered diamanes with lonsdaleite stacking sequence (AAA…) are denoted as D(AA) and D(AAA). In the left part of the Fig. 1c-d the D(AB) and D(ABC) structures are presented whereas on the right possible stacking sequences of carbon layers in corresponding diamanes are shown. In general, graphane can be named as D(A) but we suppose that graphane is the intermediate structure between graphene and diamane due to the structure does not contain any diamond like cells.

Formation energy (eV/atom) of studied structures was calculated according to the following equation [21,22]: $E_{form} = (E_{str} - nE_{graphene} - 2E_{H_2})/(n+2)$, where $E_{graphene}$ is energy of graphene per carbon atom (-9.30 eV/atom), $E_{H_2}$ is energy of hydrogen molecule (-2.99 eV/atom) and $n$ is number of carbon atoms in the unit cell.

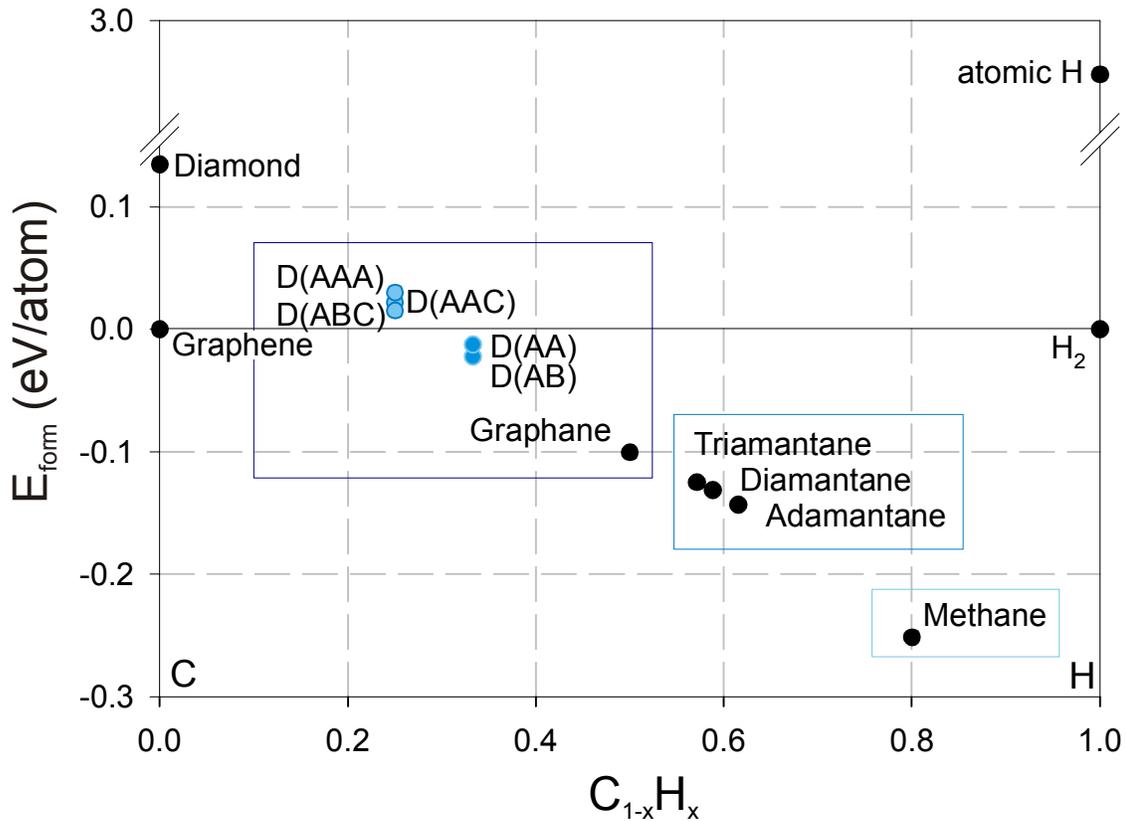

**Fig. 2. a)** Formation energy of different hydrocarbon structures as a function of hydrogen content. One can see three groups of structures marked by rectangles. The first group consists of sp³-hybridized CH₄ molecule. The next group consists of several members of diamondoid family (adamantane, diamantine and triamantane [23]). The last group consists of 2D nanoclusters of graphane and diamanes D(ij) and D(ijk).

For the sake of comparison the formation energies of a set of hydrocarbon structures were calculated using the same PBC DFT method. Studied structures can be divided on three groups. The first group (molecules) contains one member, $CH_4$ molecule which has the lowest formation energy (-0.25 eV/atom). The next group consists of diamondoids and contains three smallest members of the family: adamantane, diamantine and triamantane. [23] And the third group (two-dimensional diamond-like nanothick films) contains graphane and diamanes.

The energies of the studied structures tend nearly linear to the energy of bulk diamond (Fig. 2a) upon the hydrogen content. The diamanes with diamond type layer stacking (AB and ABC types) display the lowest energy per atom in comparison with corresponding structures with other stacking sequences but the difference in the energy of diamanes with diamond and other stacking sequences is smaller than 0.02 eV/atom which justifies the possible existence of diamanes with any stacking types. Two layered diamanes D(ij) are less favorable than graphane but are more favorable than graphene. The stability of the D(ij) is also proved by the high energy of 0.74 eV/atom of separation of D(AB) to two isolated carbon layers hydrogenated from one side.

Fig. 2a allows to estimate the possible way of preparation of diamanes. The formation energy of three-layered diamanes is positive (in the assumption of zeros temperature and pressure) if graphene and $H_2$ are used as source species of carbon and hydrogen atoms, respectively. The changing of source of hydrogen from molecular $H_2$ to atomic H leads to different sign of the formation energy from positive to negative which means that all studied diamane series become energy favorable. It should be noted that the graphane-like structure was synthesized experimentally using by atomic hydrogen.[11] In the case of diamanes, passivating hydrogen layers display only hexagonal symmetry as against graphane on which surface hydrogen can be arranged in various manners.[9,10,24]

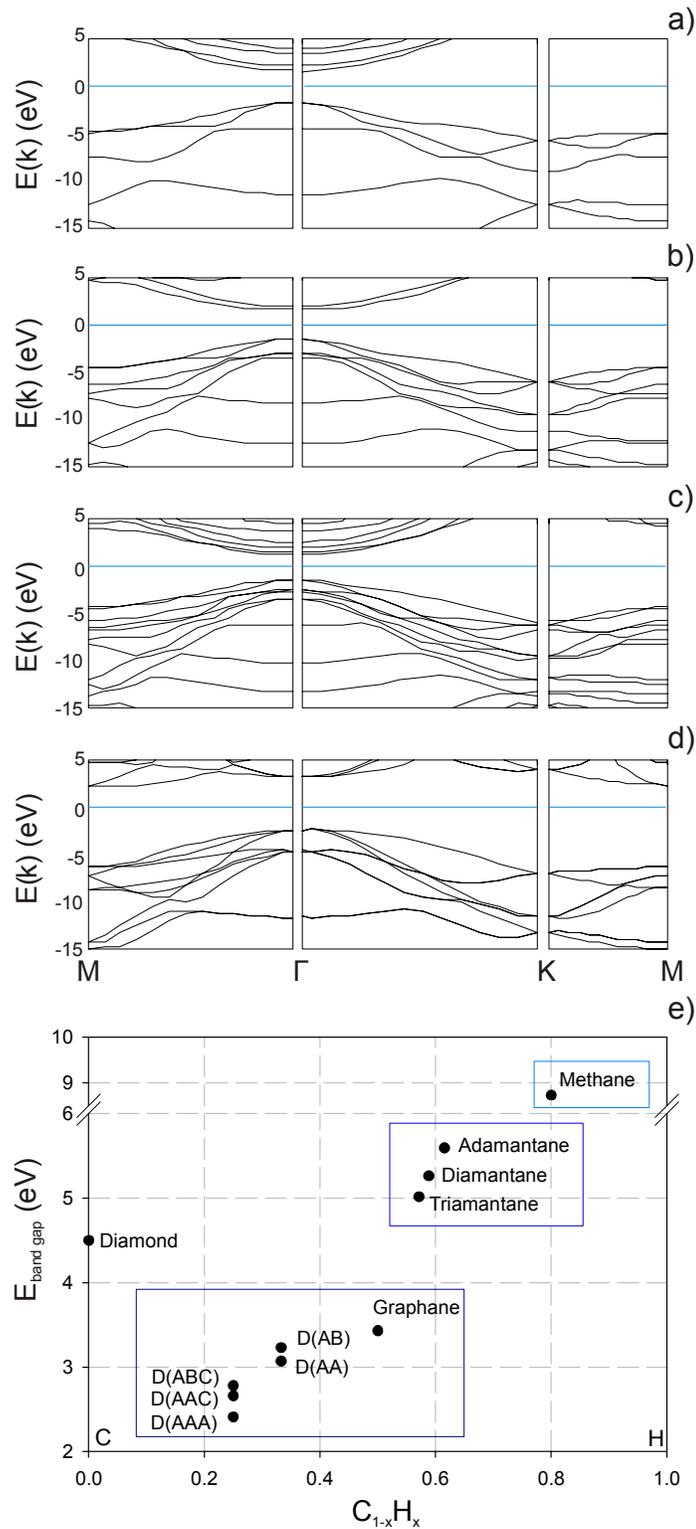

**Fig. 3.** Evolution of the band structures of 2D carbon nanostructures: a) graphane; b) D(AB); c) D(ABC) and d) in the limit of infinite number of layers (bulk diamond with the same orientation of reciprocal vectors ). Fermi level is marked by horizontal blue line; e) band gap width of different hydrocarbon structures as a function of hydrogen content. One can see three groups of structures marked by rectangles. The first group consists of $sp^3$-hybridized $CH_4$ molecule. The next group consists of several members of diamondoid family (adamantane, diamantine and triamantane [23]). The last group consists of 2D nanoclusters of graphane and diamanes D(ij) and D(ijk).

Let us consider the electronic properties of the studied structures. Band structures of graphane and diamanes are similar and display dielectric behavior and direct band gap (Fig. 3). For the sake of comparison, the electronic structure of methane and diamondoid clusters was also calculated. The band gap of studied structures should tend to the band gap of bulk diamond (Fig. 3e) upon the decrease of hydrogen content (and increase of thickness of the structure). In the case of diamondoid clusters similar results were obtain in Ref. 25. The diamane band gap widths are lower than the gap of graphane (3.4 eV) which evidence of the existence of minimum in the dependence of band gap value upon the number of layers because calculated band gap of diamond (diamane with infinite index) is 4.5 eV. The nonlinear effect of the studied films can be explained by surface states and quantum confinement effect. During the increasing of diamane film thickness the contribution of electrons from the bulk increases and electronic properties of the diamanes tends to the properties of diamond.

The phonon band structures of graphane, D(AB) and D(ABC) are presented in Fig. 4 (a, b and c, respectively). The energy splitting of the graphane highest active modes in Raman spectrum (2856 cm$^{-1}$ and 2896 cm$^{-1}$, reference values [10] are 2842 cm$^{-1}$ and 2919 cm$^{-1}$) is equal to 40 cm$^{-1}$ whereas diamanes display smaller energy splitting of the modes (2865 cm$^{-1}$ and 2875 cm$^{-1}$ for D(AB), and 2874 cm$^{-1}$ and 2882 cm$^{-1}$ for D(ABC), respectively), see group VI in Table 1. The increasing of the number of modes in the frequency region around 1332 cm$^{-1}$ (diamond fingerprints) can be an indication of diamane films due to linear increase of line intensities upon thickness of the diamond-like film [26] (group V in Table 1). Another characteristic feature of diamanes is appearance of the vibrational modes at 664 cm$^{-1}$ and 848 cm$^{-1}$ in the cases of D(AB) and D(ABC), respectively (group III in Table 1). Near 500 cm$^{-1}$ frequency region one can see two (for D(AB)) and five (for D(ABC)) optical modes bunched in one and two groups, respectively (groups I and II in Table 1).

Three acoustic branches of graphane and diamanes correspond to in-plane (two branches, linear dependence $\omega(k)$) and out-plane (one branch, quadratic dependence $\omega(k)$) vibrations of the 2D structure. [27] The increasing of the thickness of the film leads to gradual transformation of the quadratic branch to linear one of transverse mode typical for the crystal.

**Table 1. Phonon frequencies (cm$^{-1}$) at Γ-point region and velocities of transverse and longitudinal acoustic in-plane modes (10$^3$ m/s)**

| | ω$_{opt}$ groups | | | | | | v$_{TA}$, | v$_{LA}$, |
|---|---|---|---|---|---|---|---|---|
| | I | II | III | IV | V | VI | | |
| Graphane | - | - | - | 1123, 1123 | 1159, 1162, 1162, 1328, 1328 | 2856, 2896 | 12.0 | 17.7 |
| D(AB) | 401, 401 | | 664 | 1131, 1131, 1133, 1133 | 1201, 1249, 1260, 1260, 1313, 1313 | 2865, 2875 | 12.1 | 17.8 |
| D(ABC) | 291, 291 | 467, 483, 483 | 848 | 1132, 1132, 1133, 1133 | 1211, 1224, 1224, 1248, 1282, 1287, 1287, 1308, 1308 | 2874 2882 | 12.2 | 18.0 |
| Diamond (experiment [28]) | | | | | | | 12.4 | 18.3 |

Velocities of longitudinal and transverse acoustic in-plane modes (Table 1) were obtained from phonon spectra (Fig. 4). For comparison the longitudinal and transverse velocities were calculated for diamond in hexagonal orientation based on experimental value of elastic constants [28] (see Appendix for details). It is clearly seen that the velocities are gradually increased with increasing of the thickness of the films due to the augmentation of stiffness of the structures and tend to the diamond values.

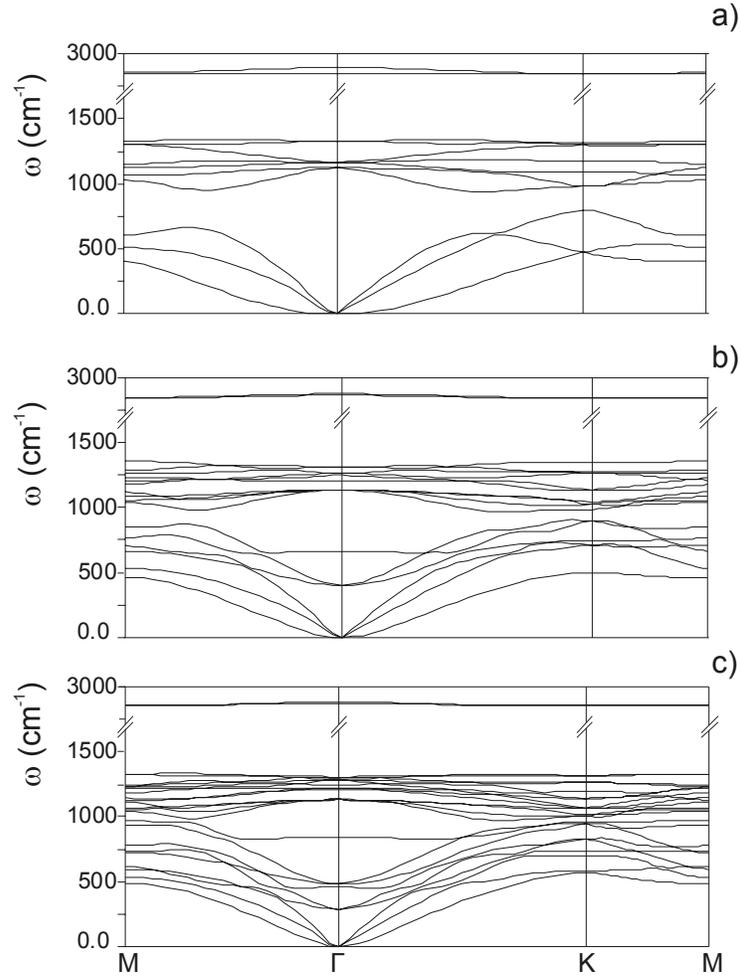

**Fig. 4. Phonon band structures of 2D carbon nanostructures: a) graphane; b) D(AB); c) D(ABC).**

Based on the velocities, elastic constants of graphene, graphane, D(AB) and D(ABC) were calculated and compared with experimental data for graphite [29] and theoretical data for graphene [30,31] and graphane. [32] Due to 2D nature of the objects under study, the elastic constants were calculated using equations $C_{11} = v_{LA}^2 \rho_{2D}$ and $C_{12} = C_{11} - 2v_{TA}^2 \rho_{2D}$, where $\rho_{2D}$ is formal density (kg/m² units) in which ambiguous thickness of 2D material is neglected. To compare the calculated data and experimental values of elastic constants of bulk graphite from Ref. 29, the theoretical data of graphene from Ref. 31 were multiplied by formal value 3.35 Å (the distance between graphene layers in graphite).

Table 2. Density $\rho_{2D}$, elastic constants $C_{11}$ and $C_{12}$, Poisson's ratio $\sigma$ of 2D nanostructures

| | $\rho_{2D}$ ($10^{-7}$ kg/m$^2$) | $C_{11}$ (N/m) | $C_{12}$ (N/m) | $\sigma$ |
|---|---|---|---|---|
| Graphene | 7.55 | 349, 358.1,[30] 308.2,[31] 355.1±20.1 [29] | 61.5, 60.4,[30] 80.4,[31] 60.3±6.7 [29] | 0.176, 0.167,[30] 0.261,[31] 0.16,[32] 0.17±0.01 [29] |
| Graphane | 7.73 | 242, 243 [32], 245 [33] | 19.5 | 0.081, 0.07 [32] |
| D(AB) | 14.9 | 474 | 35.9 | 0.076 |
| D(ABC) | 22.2 | 718 | 58.3 | 0.081 |

Graphane displays lower stiffness than graphene due to the $sp^3$ corrugation of graphane structure which allows higher elasticity. [32,33] Two layered diamane has expected bigger $C_{11}$ constant than graphene but smaller $C_{12}$ which leads to smaller Poisson's coefficient (Table 2). The same result was obtained for graphane in the work of Ref. 32. In diamanes the surface and bulk carbon atoms make practically the same contribution to the stiffness of the film due to absence of surface reconstruction.

Finally we discuss the possible methods of diamane fabrication. The adsorption of hydrogen atoms on the outer layers of multilayered graphene transforms the $sp^2$ carbon bonds to $sp^3$ ones. Such layers display deficiency of electrons and can connect with inner layers of the structure and form diamond cells and eventually diamane film.

Previously [13] it was shown that during the formation of bonds between bigraphene layers with adsorbed hydrogen atoms the formation energy goes down, demonstrating a possibility of diamane synthesis. One more possibility to synthesize diamanes is an overlapping of the ends of graphene ribbons or graphenes (Fig. 5). The Brenner empirical potential [34] was used for the relaxation and calculation of energy stability of two overlapped ribbons. It was obtained that such structure is stable and energetically favorable.

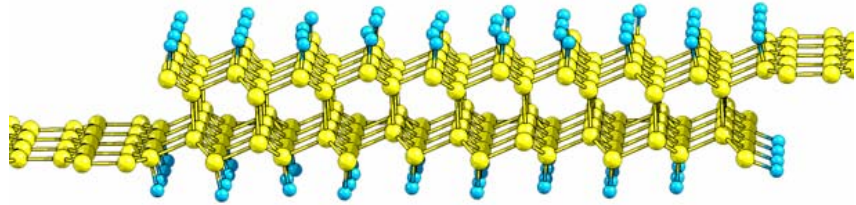

Fig. 5. The relaxed structure of two graphene ribbons with diamane region between them.

Hydrogenated graphene layers were studied in the paper of Luo *et al.* [35] and Raman spectra of hydrogenated single, double, triple etc. graphene were obtained. The increasing of intensity of D band in the vicinity of 1340 cm$^{-1}$ region with increasing of the number of graphene layers can be associated with appearance of diamane parts in the structure. One more experimental evidence of diamane growth is a synthesis of diamond films with crystalline structure. [36]

The electronic structure calculations directly demonstrate complex dependence of the physical properties of diamanes upon their thickness. Controllable variation of the number of $\langle 111 \rangle$ layers and sequence order leads to tunable variation of electronic properties of diamanes. Diamanes can be applied in nanoelectronics and nanooptics e.g. as active laser medium within lasers. The increasing number of layers in diamane will lead to transition of direct band gap to indirect one (diamond) therefore only thin films can be used for this purpose. Diamanes can be used as optical planar waveguides and as very thin dielectric hard films in nanocapasitors or as chemical nanosensors (e.g. it was obtained that hydrogenated diamond surface changes their electrical conductivity by adsorption of $H_3O^+$ species [37]) or as mechanically stiff nanothick elements in nanoelectronics.

*Acknowledgements*

L.A.C. was supported by the Russian Academy of Sciences, program No. 21 and by the Russian Foundation for Basic Research (project no. 08-02-01096). P.B.S. and B.I.Y. acknowledge support by the Office of Naval Research (MURI project). P.V.A. and P.B.S. also acknowledge the collaborative RFBR-JSPS grant No. 09-02-92107-ЯФ. We are grateful to the Joint Supercomputer Center of the Russian Academy of Sciences for the possibility of using a cluster computer for quantum chemical calculations. The geometry of all presented structures was visualized by commercial ChemCraft software

*Appendix*

In the case of hexagonal diamane films longitudinal and transverse velocities can be measured in the X ($\langle 21 \cdot 0 \rangle$) direction which corresponds to $\langle 112 \rangle$ directions of cubic diamond. Solution the Christoffel's equation of bulk acoustic wave propagation for cubic crystals in the $\langle 112 \rangle$ direction gives us three eigenvalues $\lambda_i = \rho_0 v_i^2$ ($i = 1,2,3$) associated with pure shear (S), quasi-shear (QS) and quasi-longitudinal (QL) wave velocities $v_i$, respectively (see the Table). Note that phonon spectroscopy is commonly used an appropriate notation of these modes as TA and LA phonons.

**Table 3.** The solution of Christoffel's equation for $\langle 112 \rangle$ direction of cubic diamond

| $\vec{N}$ | $\vec{U}$ | Wave | $\lambda_i = \rho_0 v_i^2$ |
|---|---|---|---|
| [112] | $[1\bar{1}0]$ | S | $\lambda_1 = \frac{1}{6}(C_{11} - C_{12} + 4C_{44})$ |
| [112] | $\sim[11\bar{1}]$ | QS | $\lambda_2 = \frac{1}{2}(\Gamma_{11} + \Gamma_{12} + \Gamma_{33}) - \frac{1}{2}\sqrt{(\Gamma_{11} + \Gamma_{12} - \Gamma_{33})^2 + 8(\Gamma_{13})^2}$ |
| [112] | $\sim[112]$ | QL | $\lambda_3 = \frac{1}{2}(\Gamma_{11} + \Gamma_{12} + \Gamma_{33}) + \frac{1}{2}\sqrt{(\Gamma_{11} + \Gamma_{12} - \Gamma_{33})^2 + 8(\Gamma_{13})^2}$ |
| $\Gamma_{11} = \frac{1}{6}(C_{11} + 5C_{44})$ $\quad$ $\Gamma_{12} = \frac{1}{6}(C_{12} + C_{44})$ $\quad$ $\Gamma_{13} = \frac{1}{3}(C_{12} + C_{44})$ $\quad$ $\Gamma_{33} = \frac{1}{3}(2C_{11} + C_{44})$ ||||

Here $\vec{N}$ and $\vec{U}$ are the unit vectors of wave front and particles' displacement, respectively; $\rho_0$ is the bulk crystal density of diamond ($\rho_0 = 3.515$ g/cm$^3$); $C_{ij}$ and $\Gamma_{ij}$ are the components of elastic stiffness constants and Christoffel's tensor, respectively.

It is evident that second mode is absent in the case of 2D crystal as far as its particles' displacement is associated with off-plane vibration. On the contrary, first and third modes are associated with in-plane transverse and longitudinal phonons. So the first and third modes were used to obtain the transverse and longitudinal wave velocities in the $\langle 21\cdot 0 \rangle$ diamane's direction by $C_{ij}$ experimental constants of bulk diamond. [28]